\documentclass{article}

\usepackage[capitalize]{cleveref}
\usepackage{xspace}
\usepackage{url}
\usepackage{graphicx}
\usepackage{verbatim}

\newcommand{\ci}{CI/CD\xspace}
\newcommand{\st}{2012\xspace}

\author{%
Hugo da Gião\textsuperscript{1,2}, 
André Flores\textsuperscript{1,2}\\ 
Rui Pereira\textsuperscript{3}, 
Jácome Cunha\textsuperscript{1,2}\\
	hugo.a.giao@inesctec.pt, up201907001@edu.fe.up.pt\\ rui.alexandre.pereira@outsystems.com, jacome@fe.up.pt}

\date{
\footnotesize
\textsuperscript{\textbf{1}} Faculty of Engineering, University of Porto, Portugal\\ 
\textsuperscript{\textbf{2}} INESC TEC, Portugal\\
\textsuperscript{\textbf{3}} OutSystems, Portugal}

\title{Chronicles of CI/CD: A Deep Dive into its Usage Over Time}

\begin{document}

\maketitle



\begin{abstract}
DevOps is a combination of methodologies and tools that improves the software development, build, deployment, and monitoring processes by shortening its lifecycle and improving software quality. Part of this process is \ci, which embodies mostly the first parts, right up to the deployment. Despite the many benefits of DevOps and \ci, it still presents many challenges promoted by the tremendous proliferation of different tools, languages, and syntaxes, which makes the field quite challenging to learn and keep up to date.

Software repositories contain data regarding various software practices, tools, and uses. This data can help gather multiple insights that inform technical and academic decision-making. GitHub is currently the most popular software hosting platform and provides a search API that lets users query its repositories.

Our goal with this paper is to gain insights into the technologies developers use for CI/CD by analyzing GitHub repositories. 
%We collect various tools from surveys, reports, and blog posts. 
Using a list of the state-of-the-art \ci technologies, we use the GitHub search API to find repositories using each of these technologies. We also use the API to extract various insights regarding those repositories. We then organize and analyze the data collected.

From our analysis, we provide an overview of the use of \ci technologies in our days, but also what happened in the last 12 years. We also show developers use several technologies simultaneously in the same project and that the change between technologies is quite common. From these insights, we find several research paths, from how to support the use of multiple technologies, both in terms of techniques, but also in terms of human-computer interaction, to aiding developers in evolving their \ci pipelines, again considering the various dimensions of the problem.

%In our analysis, we found several insights concerning the tools studied. According to our analysis, these insights include that Docker is the most popular tool. We also found that the tools used for the Deployment phase are some of the most popular. We also found that, on average, repositories associated with more popular tools had lower interaction.
 
\end{abstract}


\section{Introduction} \label{sec:introduction}

In 2022 the Information Technology (IT) market accounted for \$9,325.69 billion dollars worldwide~\cite{information2022}. Half of IT costs are related to the maintenance and upkeep of existing systems, and of those, 50\% correspond to emergencies, unplanned work, and changes~\cite{kim2016devops}. This cost happens partially because the IT and development team's goals often conflict in most organizations. Instead of working for a common goal, most organizations' development teams are responsible for reacting to and answering rapid changes in markets and customer needs. In contrast, IT teams provide customers with a secure, reliable, and stable experience. This dynamic leads to technical debt, meaning decisions made over time lead to increasingly more complex problems and slower operations and thus impede the achievement of the organization's goals~\cite{kim2016devops}.

In recent years, DevOps, that is, the integration of development and IT teams seamlessly, has emerged to enable organizations to respond to market demands with unparalleled agility hoping to address the aforementioned problems. In particular, methodologies and technologies for Continuous Integration, Continuous Delivery, and Continuous Deployment, or simply CI/CD, have emerged as a means for organizations to achieve rapid and frequent delivery of changes. A CI/CD pipeline encompasses a series of essential steps involved in integrating and deploying codebase changes~\cite{kim2016devops}. This process involves the regular integration of new code changes, which undergo automated building and testing procedures. Subsequently, the validated code is deployed into production through the CD process~\cite{kim2016devops}.
%\end{sloppypar}

Organization have the option to employ various technologies like GitHub Actions\footnote{\url{https://github.com/features/actions}}, GitLab CI/CD\footnote{\url{https://docs.gitlab.com/ee/ci/}}, Travis CI\footnote{\url{https://www.travis-ci.com/}}, CircleCI\footnote{\url{https://www.jetbrains.com/teamcity}} or Jenkins\footnote{\url{https://www.jenkins.io/}}. The wide range of existing technologies makes it harder for researchers and developers to study and improve the CI/CD process. Indeed, there is no view of the usage of \ci technologies in a global sense, e.g., most used technologies, their evolution over time, which kinds of projects most rely on \ci, and so on.

GitHub is a widely used version control and software hosting service. As of 2023, more than 100 million developers used GitHub, and the platform hosted over 284  million public repositories~\cite{octoverse}. GitHub also provides an API to collect information from its software repositories. With this API, we can search repositories inside GitHub using parameters such as keywords in their name and  README, size,  number of stars, followers, and forks\footnote{\url{https://docs.github.com/en/rest/search}}. The insights collected using the API help understand various developer behaviors and have been used in several studies~\cite{challengesdocker,Henkell,golzadeh_rise_2022,cicd_open_src_android,mlops_github_pratices}.
Indeed, in this work, we build on GitHub to gain insights into how developers use several \ci technologies. The information collected can guide future research on possible improvements to the DevOps process and in particular for \ci. With this work, we thus aim to answer the following research questions:

\begin{description}

\item[RQ1] What characterizes the current landscape of CI/CD?

In addressing Research Question 1 (RQ1), our objective is to gain a comprehensive understanding of the prevailing utilization of diverse CI/CD technologies. Our aim is to identify patterns that can serve as catalysts for future (research) endeavors, both within industrial and academic contexts.

From our sample of repositories, containing 612557 individuals, we know that about 32.7\%
%200023/612557 = 32.7
include some \ci technology. In \cref{sec:rq1} we explore the most used technologies and their relationship with programming languages.
Indeed, projects in some languages do not rely often on \ci which deserves further investigation and possibly better solutions for those developers.

\item[RQ2] Can a single CI/CD technology adequately meet the needs of a specific project?

With RQ2, our goal is to understand to which extent the technologies can serve the projects' needs.

Indeed, we have discovered that  30271 projects use at least two technologies and that some (few cases) use up to 13 different technologies at the same time.
This motivates further investigation on the interoperability of the different technologies and the support developers have to work with multiple \ci technologies at the same time.

\item[RQ3] What is the evolution of CI/CD usage over time?

With this research question, our goal is to find how projects change over time regarding the \ci technologies used. Our goal is also to get an overview of changes in pipelines over time.

Over the years, there have been two major technologies used. Moreover, it is clear that developers often change technologies. Since currently there is little support for these evolutions, this paves a very interesting path for future research.

\end{description}

We organize the rest of the document  as follows:
\cref{sec:related} presents several works related to our own. Those works focus on mining information about several DevOps aspects in various software repositories.
In \cref{sec:methodology}, we present the methodology used for this work. This methodology includes how we found the \ci technologies for this study, what information we have collected from the repositories and how it is related to our research questions, how we used the GitHub API to collect information from the repositories, and how we organized the data collected from the said repositories. 
In Sections~\ref{sec:rq1},~\ref{sec:rq2}, and ~\ref{sec:rq3}, we showcase and analyze the data for each research question.
In \cref{sec:discussion} we answer our RQs and provide several future research paths based of this work.
\cref{sec:threats} discusses the threats to the validity of our work.
Finally, \cref{sec:conclusions} presents our conclusions and future work.

\section{Related work} \label{sec:related}

In this section, we introduce our related work. This paper's related work focuses on several articles that mine software repositories to understand and improve aspects of software development. Similarly to these works, we mine repositories related to the \ci. 
%However, in our work, we do not search for specific scripts and identify the repositories using keywords in their title, description, and README. We also do not analyze specific files but study the repositories using the GitHub search API data. We also analyze repositories that contain several technologies with different purposes and uses.

Xu et al. \cite{xu} introduce the idea of mining container image repositories for configuration and other deployment information of software systems. The authors also showcase the opportunities based on concrete software engineering tasks that can benefit from mining image repositories. They also summarize the challenges of analyzing image repositories and the approaches that can address these challenges. 
These authors focus their work on technologies for deployment, while with our work we give a broader overview of the usage of \ci technologies.

Wu et al. \cite{wu} present a preliminary study on 857,086 Docker builds from 3,828 open-source projects hosted on GitHub. Using the Docker build data, the authors measure the frequency of broken builds and report their fix times. They also explore the evolution of Docker build failures across time. This work is focused on a particular technology, whilst ours covers many.

Zahedi et al. \cite{Zahedi} present an empirical study exploring continuous software engineering from the practitioners' perspective by mining discussions from Q\&A websites. The authors analyzed 12,989 questions and answers posted on Stack Overflow. The authors then used topic modeling to derive the dominant topics in this domain. They then identify and discuss key challenges. Although the studied topic is related to \ci, we analyzed concrete software projects.

Mazrae et al. \cite{rostami_mazrae_usage_2023} present a qualitative study of CI/CD technologies usage, co-usage, and migration based on in-depth interviews. They identify reasons for the use of specific technologies, reasons for co-usage of CI/CD technologies in the same project, and migrations executed by the interviewees. Their study reveals a clear trend in migration from Travis to GitHub Actions. We have used a very different source of data, but some of the conclusions of our study are in line with Mazrae et al.

Goldazeh et al. \cite{golzadeh_rise_2022} conduct a qualitative analysis of the usage of seven popular CI technologies in the GitHub repositories of 91m810 active npm packages having used at least one CI service over a period of nine years. Their findings include the fall of Travis, the rapid rise of GitHub Actions, and the co-usage of multiple CI technologies. These results are in line with ours, but we have investigated many more projects and \ci technologies.

Liu et al. \cite{cicd_open_src_android} mine 84,475 open-source Android applications from GitHub, Bitbucket, and GitLab to search for \ci adoption. They find only around 10\% applications leverage \ci technologies, a small number of applications (291) adopt multiple \ci technologies, nearly half of the applications that adopt \ci technologies do not really use them, and \ci technologies are useful to improve project popularity. Their approach is similar to ours however, our analysis is done with a greater sample of 612557 repositories and we don't limit ourselves to Android applications.

Calefato et al. \cite{mlops_github_pratices} study MLOps (that is, DevOps but focused on machine learning projects) practices in GitHub repositories, focusing on GitHub Actions and CML. Their preliminary results suggest that the adoption of MLOps workflows is rather limited. On the other hand, we have found that, indeed, many projects rely on \ci pipelines.

Kumar et al. \cite{helena2_devops_practices} assess the maturity of DevOps practices in the software industry. To this end, they analyze the HELENA2 dataset (an international survey aiming to collect data regarding the common use of software and systems in practice). They rank organizations by DevOps maturity. The authors conducted a user survey while we analyzed software repositories.

\section{Data collection} \label{sec:methodology}

\begin{figure}[!ht]
\centerline{\includegraphics[width=.8\columnwidth]{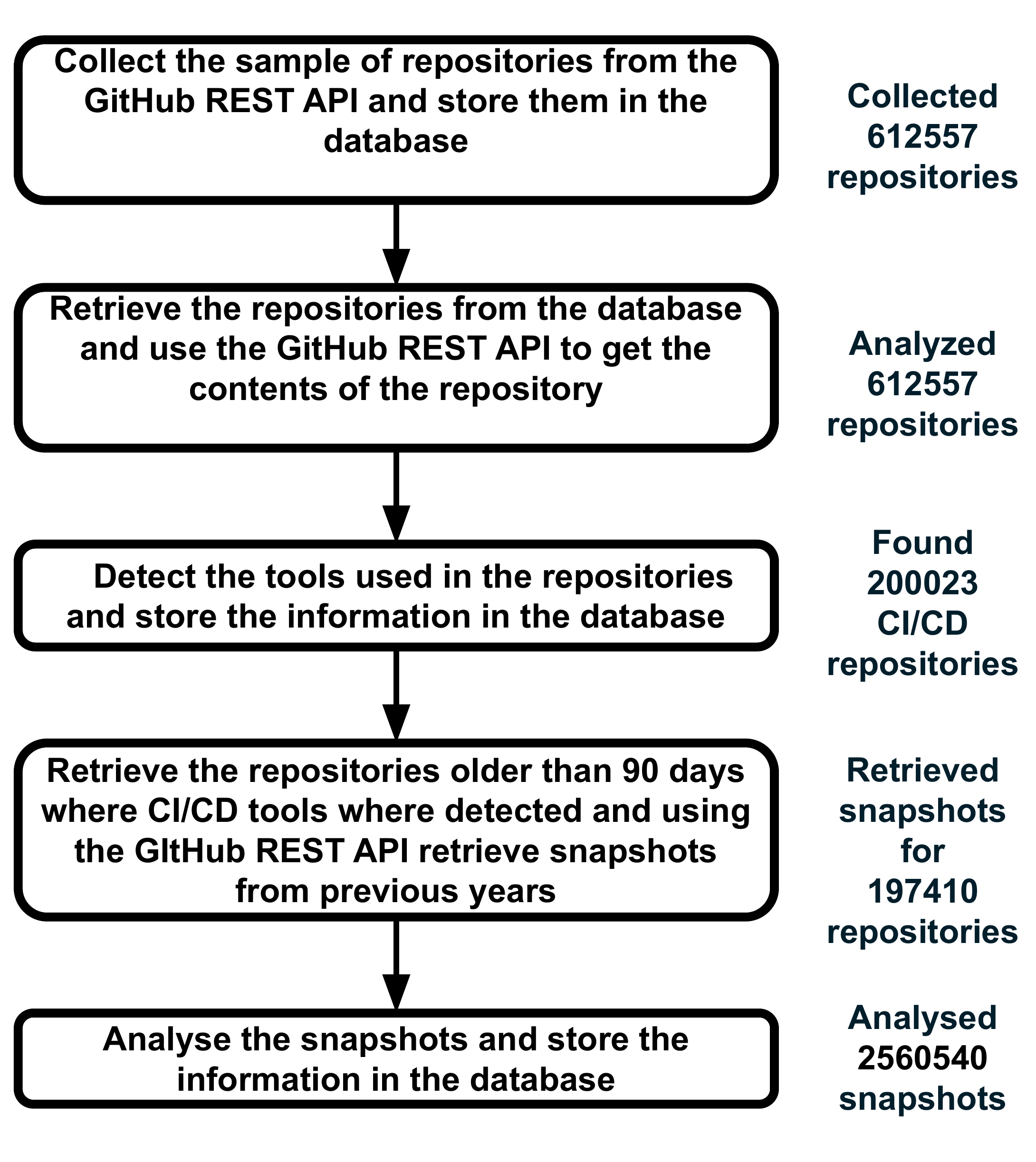}}
\caption{Data collection process.}
\label{fig:process}
\end{figure}

This section details the process we followed to collect the data for our analysis. \cref{fig:process} presents an overall view that we detail in the following paragraphs.

\subsection{Repository Sampling}

The initial data collection phase involved assembling a representative sample of GitHub repositories, reflective of real-world projects. To achieve this, we excluded school projects and smaller repositories, employing a methodology inspired by prior research projects. That is, we focused on repositories attaining a certain level of popularity, ultimately choosing to collect only those with more than ten stars, accounting for a mere 5\% of all GitHub repositories.

The sampling process encompassed repositories from the beginning of 2012 to October 18, 2023. This timeframe was selected because the concept of DevOps gained prominence around that period. This assertion is supported by the emergence of the first comprehensive survey on the state of DevOps by Puppet Labs in the same year~\cite{PuppetbyPerforce}. %Additionally, most tools integral to our investigation were released after 2012.

Leveraging the GitHub REST API, we devised a methodology to identify the top one thousand repositories with the highest star count each week. That is, we queried GitHub and got one thousand repositories every week in the time range. From these, we excluded the ones with fewer than ten stars. This approach yielded a comprehensive dataset, totaling 612,557 repositories, spanning the specified date range. The entire dataset of repositories is provided for reference \cite{Author2023}. The code used to create this dataset is also available \cite{coderq1rq2}.

\subsection{\ci Technologies}

Following the collection phase, our objective was to discern the technologies employed in each repository. We used a dataset of 61 CI/CD technologies curated by the Cloud Native Computing Foundation \cite{Study_2023}, established by the Linux Foundation in 2015.
%, we categorized the tools into distinct groups. This classification included cloud tools that evade detection due to a lack of artifacts in repositories in \cref{table:cloud}, deprecated tools lacking current documentation in \cref{table:deprecated}, libraries introducing unnecessary complexity in \cref{table:libraries}, and tools meriting the criteria for our further analysis in \cref{table:used}. 
%
We identified pertinent artifacts and patterns for each technology in the dataset, enabling us to automatically recognize repositories utilizing those technologies. This process involved analyzing file trees accessed through the GitHub REST API and scrutinizing file contents with extensions compatible with the respective technologies.
Each technology was identified using one of two heuristics: \textit{i)} some technologies use files with a particular extension; \textit{ii)} for the others, we inspected specific types of files (e.g., YAML files) searching for content specific to the underlying technology. From this process, we divided the initial 61 technologies into different categories: \textit{i)} 39 technologies that we could identify (\cref{table:used}); \textit{ii)} 10 which we could not identify since there was not a clear artifact that we could use (\cref{table:notused}); \textit{iii)} 4 technologies that required the use of specific code to be identified as they are libraries embedded in the code (\cref{table:libraries}); and \textit{iv)} 1 deprecated technology with no current documentation that made it impossible to recognize (\cref{table:deprecated}).

\begin{table}[!htb]
\caption{\ci technologies we can identify and analyze.}
\centering
\begin{tabular}{ l |l }  
  \hline
Agola & AppVeyor\\
\hline
ArgoCD & Bytebase\\
\hline
Cartographer & CircleCI\\
\hline
Cloud 66 Skycap & Cloudbees Codeship\\
\hline
Devtron & Flipt\\
\hline
GitLab & Google Cloud Build\\
\hline
Helmwave & Travis\\
\hline
Jenkins & JenkinsX\\
\hline
Keptn & Liquibase\\
\hline
Mergify & OctopusDeploy\\
\hline
OpenKruise & OpsMx\\
\hline
Ortelius & Screwdriver\\
\hline
Semaphore & TeamCity\\
\hline
werf & Woodpecker CI\\
\hline
GitHubActions & Codefresh\\
\hline
XL Deploy & Drone\\
\hline
Flagger & Harness.io\\
\hline
Flux & GoCD\\
\hline
Concourse & Kubernetes\\
\hline
AWS CodePipeline & \\
\hline
\end{tabular}
\label{table:used}
\end{table}

\begin{table}[!htb]
\caption{\ci technologies that we cannot identify due to the lack of clearly identifiable artifacts.}
\centering
\begin{tabular}{ l |l }  
  \hline
Akuity  & Bamboo\\
\hline
Buildkite  & Bunnyshell\\
\hline
CAEPE  & Keploy\\
\hline
Northflank  & OpenGitOps\\
\hline
Ozone  & Spacelift\\
\hline
\end{tabular}
\label{table:notused}
\end{table}

\begin{table}[!htb]
\caption{Libraries introducing unnecessary complexity.}
\centering
\begin{tabular}{ l |l }  
  \hline
Brigade  & k6\\
\hline
OpenFeature  & Unleash\\
\hline
\end{tabular}
\label{table:libraries}
\end{table}

\begin{table}[!htb]
\caption{Deprecated technologies lacking documentation.}
\centering
\begin{tabular}{ l }  
  \hline
D2iQ Dispatch  \\
\hline
\end{tabular}
\label{table:deprecated}
\end{table}

\subsection{The Repositories with \ci Technology}

Our analysis focused on the sample acquired earlier, specifically the contents of the latest commit to the main branch of each repository. To ensure reproducibility, we recorded the SHA of the file tree (available in the database of our dataset). A total of 200,023 repositories were identified as utilizing one or more of the \ci technologies identified in the previous section. The comprehensive dataset containing the repositories and the corresponding technologies is also made available for further examination \cite{datasettools}.  We use this dataset to analyze the state of \ci according to the most recent snapshots. The code used to create this dataset and the figures to support our answers to the research questions are also available~\cite{coderq1rq2}.

To examine the past state of \ci, we retrieved snapshots of repositories over time and ran the same CI/CD technologies analysis on each snapshot. The comprehensive dataset containing the repositories and the corresponding snapshots and technologies is also made available for further examination \cite{datasettoolhisoty}. The code used to create this dataset and the figures to support our answers to the research question are also available \cite{reporq3}.

We ran a test on a random sample of 85 repositories to choose the time interval for the analysis. Snapshots were retrieved at 90-day, 180-day, and 365-day intervals for each repository. For each snapshot, we then ran CI/CD technologies analysis. Our goal was to determine the number of changes in the CI/CD tech stack we would lose by increasing the retrieval interval, a change being any difference in the CI/CD stack compared to the previous snapshot. For the sample, we found 270 stack changes at a 90-day sampling interval, 217 changes at a 180-day sampling interval, and 179 changes at a 365-day sampling interval. From a 90-day to a 180-day rate, there was a 19.6\% decrease in the detected changes, and from a 180-day to a 365-day rate, there was a 17.5\% decrease in detected changes. A lower sampling interval, or retrieving all commits for each repository, was not feasible due to GitHub API rate limits and the time for the study. Based on these results, we determined to run our analysis at a 90-day sampling rate.

A subset of the previously selected repositories was analyzed. The inclusion criteria were repositories where CI/CD technologies were detected in the latest commit and created before July 16, 2023 (so they were at least 90 days old). This resulted in a subset containing 197504 repositories.

For each selected repository, the first commit was retrieved. A sequential iterative process was employed, where the latest commit (if one existed) was retrieved for each 90-day interval starting from either January 1, 2012 or the repository's first commit date, whichever was later. This process continued until the date of the repository's last update at the time of retrieval. All commits were retrieved from the default branch of the repository. The git trees of the snapshot commits were then examined for CI/CD technology presence using the previously described methodology.

After all repositories were processed, we cleaned the retrieved data. Any snapshots from before January 1, 2012, were discarded, and the last snapshot of each repository was set to the one used in the previous analysis. For some snapshots, the GitHub API could not return a git tree. We attempted to process these snapshots again to eliminate any momentary API malfunction. The original detection method for GitHub Actions could lead to false detection when other technologies were present in the snapshot, so we reprocessed all snapshots that had more than one technology detected and GitHub Actions present. Lastly, we checked each snapshot's date and detected technologies against the detected technologies' launch dates and removed snapshots where a technology was detected before it was launched. If, at the end of these cleaning steps, a repository was left without snapshots, it was discarded.

From an initial 197504 repositories selected for temporal analysis, we retrieved the \ci technology use history of 197410. For the 94 repositories whose \ci technology history could not be retrieved, the reasons are as follows: the 19 repositories could not be processed in the initial snapshot retrieval because they had either been deleted or gone private, in the data cleaning step, another 39 repositories could not be processed because they had either been deleted or gone private, and 36 were discarded because they had no snapshots at the end of the cleaning steps.

\section{RQ1: What characterizes the current landscape of CI/CD?} \label{sec:rq1}

By answering this RQ we intend to provide a clear and accurate overview of the current usage of CI/CD technologies in the open-source realm.

\subsection*{\ci usage}

We collected 612,557 created between 2012 and 2023, that is, between the beginning of the establishment of DevOps and now. From these, 200,023, 32.7\%, contain at least one \ci technology. This means about one third of the repositories in the last twelve years have, or had, some \ci support.

\begin{figure}[!ht]
\centerline{\includegraphics[width=\columnwidth]{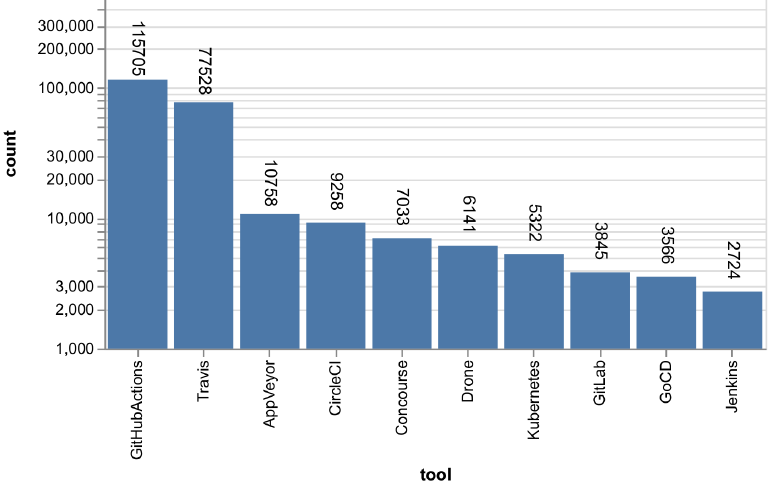}}
\caption{Distribution of technologies across CI/CD repositories (in a logarithmic scale).}
\label{fig:tools}
\end{figure}

In \cref{fig:tools}, we depict the distribution of technologies within CI/CD projects, exclusively focusing on technologies whose usage exceeds 1\% of the total repositories containing CI/CD technologies. In the online appendix of this paper\footnote{https://sites.google.com/view/msr2024}, we present a chart with all the technologies. In fact, we present several other charts in this appendix. The figure indicates that GitHub Actions significantly dominates the landscape, being present in more than half (57.8\%) 
%115705/200023
of the repositories, with Travis trailing as the second most utilized technology (38.8\%). In comparison, the majority of other technologies exhibit significantly lower usage rates in contrast to these leading two. Notably, among the 39 technologies analyzed, only ten surpass the 1\% usage threshold across CI/CD repositories.

\begin{comment}
        "values": [{"tool": "Jenkins" , "count": 2724},{"tool": "GoCD" , "count": 3566},{"tool": "GitLab" , "count": 3845},{"tool": "Kubernetes" , "count": 5322},{"tool": "Drone" , "count": 6141},{"tool": "Concourse" , "count": 7033},{"tool": "CircleCI" , "count": 9258},{"tool": "AppVeyor" , "count": 10758},{"tool": "Travis" , "count": 77528},{"tool": "GitHubActions" , "count": 115705}]

\end{comment}

The predominance of GitHub Actions seems to indicate that repository platforms have quite some influence, at least in the projects they hold. Thus, a possible conclusion is that, when considering the usage of \ci technology, developers also consider where the code will be managed.

\subsection*{\ci and programming languages}

\begin{figure}[!ht]
\centerline{\includegraphics[width=\columnwidth]{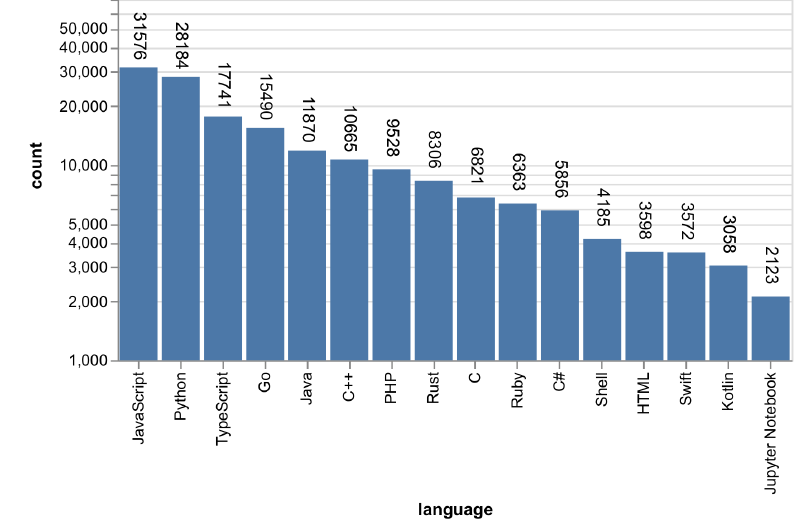}}
\caption{Programming languages in repositories containing CI/CD (logarithmic scale).}
\label{fig:pl}
\end{figure}

\cref{fig:pl} illustrates the number of repositories of each programming language with a usage rate exceeding 1\% in repositories employing at least one CI/CD technology. We use the GitHub API to get the main programming language of each repository (which does not mean the project does not use other languages). Notably, based on this depiction, JavaScript and Python emerge as the most extensively utilized languages. In absolute terms, these languages are the ones where more projects include \ci technologies. A comparison of the languages highlighted in the IEEE Spectrum's top programming languages of 2023 \cite{toplanguages} reveals noteworthy disparities. While SQL and R claim top positions in IEEE Spectrum's ranking, they are absent from our list. This discrepancy may be attributed to SQL often being embedded within other programming languages and R being predominantly utilized by less technically inclined users, with less emphasis on system development. Additionally, Jupyter Notebooks appear on our list, albeit not reaching that popularity compared to IEEE Spectrum's rankings.

The fact that projects using languages such as R, a language so popular among users without a computer science (CS) background (e.g., researchers, mathematicians), do not usually include \ci technologies, seems to indicate \ci is of difficult adoption by users without a technical (CS) background.

\begin{figure}[!ht]
\centerline{\includegraphics[width=\columnwidth]{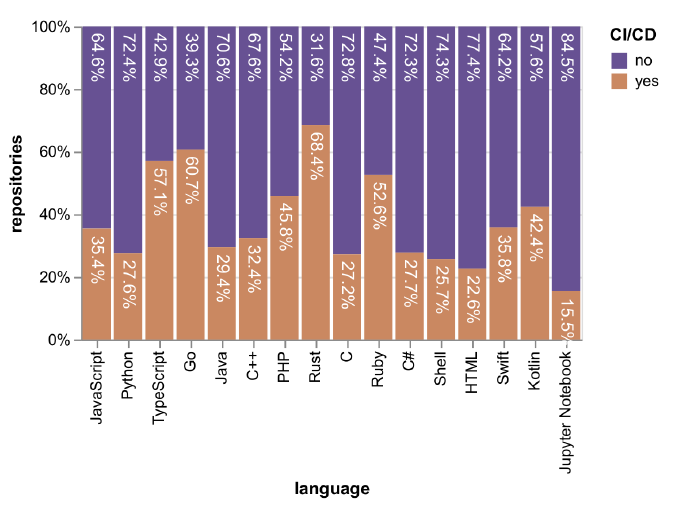}}
\caption{Percentage of CI/CD usage per language.}
\label{fig:percent}
\end{figure}

\cref{fig:percent} displays the prevalence of languages in CI/CD repositories, highlighting the percentage of sample repositories utilizing each language with and without CI/CD technologies. The figure indicates that languages like Go, Rust, and TypeScript exhibit the highest prevalence of CI/CD use within their repositories. Notably, despite Python and JavaScript having a substantial presence in CI/CD repositories, as illustrated in \cref{fig:pl}, they show a lower overall percentage of repositories utilizing CI/CD technologies.

This data seems to indicate more recent languages (e.g., Go, Kotlin, Rust, TypeScript) tend to include \ci in their projects, in some cases having more than 60\% of all projects some technology. On the other hand, more classic languages (e.g., C, C\#, HTML, Shell) seem to have less support from \ci technologies. From this, a possible conclusion is that \ci technologies find a better fit in modern languages and not so much in pre-existing ones.

\begin{figure}[!ht]
\centerline{\includegraphics[width=\columnwidth]{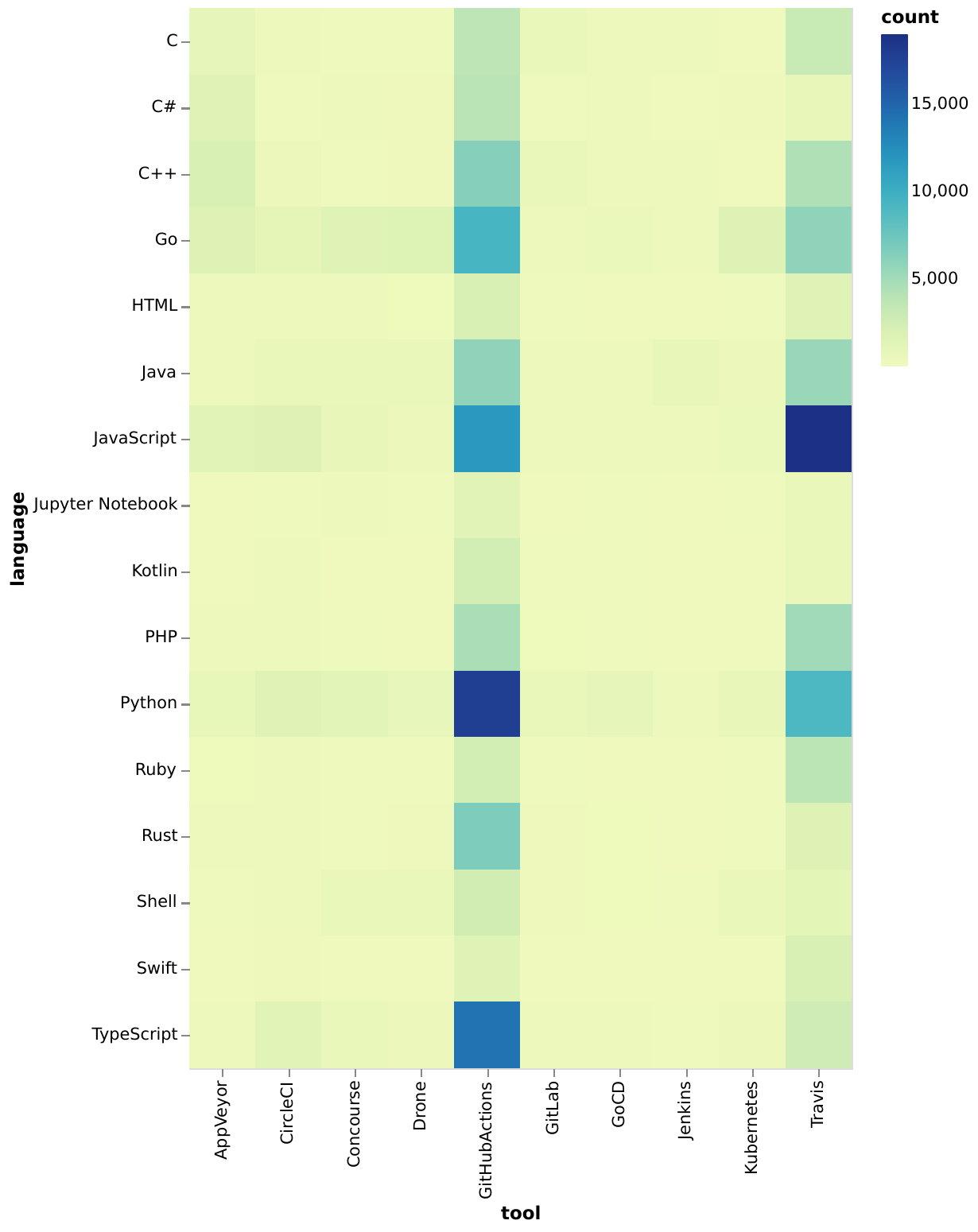}}
\caption{Heat map of technology use per programming language across CI/CD repositories.}
\label{fig:plt}
\end{figure}

\cref{fig:plt} reveals that GitHub Actions and Travis stand out as the predominant technologies utilized across various programming languages. Notably, Travis demonstrates higher popularity in JavaScript projects, while GitHub Actions is more prominently employed in Python projects. This trend may be attributed to the synchronized growth of GitHub Actions and the increasing popularity of Python. Furthermore, GitHub Actions emerges as the preferred technology across most programming languages.

This analysis suggests that programming languages experiencing heightened popularity in recent years, like Python and TypeScript, are more commonly associated with GitHub Actions. Conversely, languages such as JavaScript are more prevalent in projects utilizing Travis. This observation implies that newer projects tend to favor GitHub Actions, indicating a potential trend of migration from Travis to GitHub Actions, particularly in numerous JavaScript projects.

%\begin{comment}
% appendix
%\begin{figure}[h]
%\centerline{\includegraphics[width=\columnwidth]{rq1/graphics/maptoolslanguageslesspopular.pdf}}
%\caption{Heat map of tool use per programming language across CI/CD repositories.}
%\label{fig:plt}
%\end{figure}
%\end{comment}

\section{RQ2: Can a single CI/CD technology adequately meet the needs of a specific project?} \label{sec:rq2}

With this RQ we intend to understand the co-usage of several \ci technologies among the projects.

\begin{figure}[!ht]
\centerline{\includegraphics[width=\columnwidth]{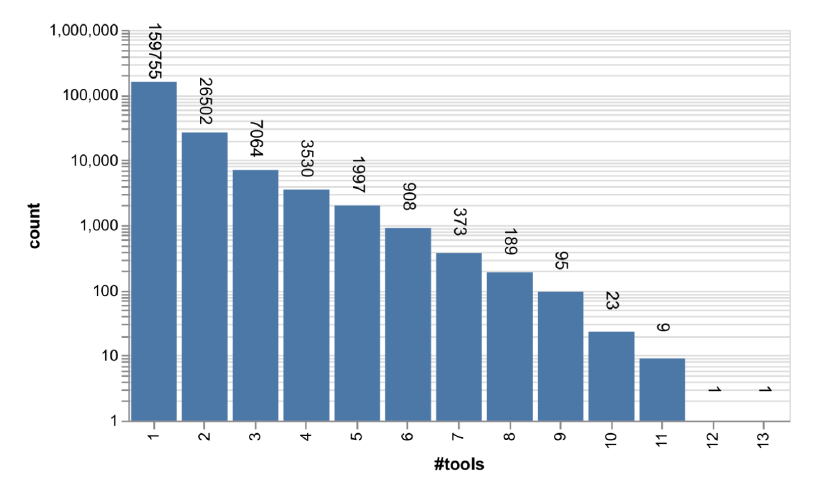}}
\caption{Distribution of the number of technologies in repositories containing CI/CD technologies (logarithmic scale).}
\label{fig:distro}
\end{figure}

\cref{fig:distro} displays the logarithmic distribution of the number of CI/CD technologies employed across repositories with at least one such technology. A notable observation from the figure is that more than three quarters of projects (84.9\%) opt for a single CI/CD technology. Nevertheless, 30271 repositories, constituting 15.1\% of all repositories with CI/CD technologies, utilize more than one technology. This suggests that while a substantial portion of projects operates effectively with a single CI/CD technology, there exists a subset of projects that necessitate the utilization of multiple technologies. This implies that certain technologies may not be adequate or suitable for specific workloads, prompting the adoption of a diverse \ci technology stack in these instances.

\begin{comment}
          {"number of tools": "1" , "count": 159755},{"number of tools": "4" , "count": 3530},{"number of tools": "2" , "count": 26502},{"number of tools": "3" , "count": 7064},{"number of tools": "5" , "count": 1997},{"number of tools": "6" , "count": 908},{"number of tools": "7" , "count": 373},{"number of tools": "9" , "count": 95},{"number of tools": "8" , "count": 189},{"number of tools": "10" , "count": 23},{"number of tools": "12" , "count": 1},{"number of tools": "11" , "count": 9},{"number of tools": "13" , "count": 1}

\end{comment}

The fact that many projects require using two or more \ci technologies raises several concerns. To the best of our knowledge, the interoperability of these technologies has not been properly studied. This deserves an investigation of its own. Moreover, \ci supporting tools (e.g., editors) are not tailored for co-usage, which makes their co-usage more difficult for developers. Thus, to study how to aid developers in these projects becomes an interesting research path.

\begin{figure}[htbp]
\centerline{\includegraphics[width=1.03\columnwidth]{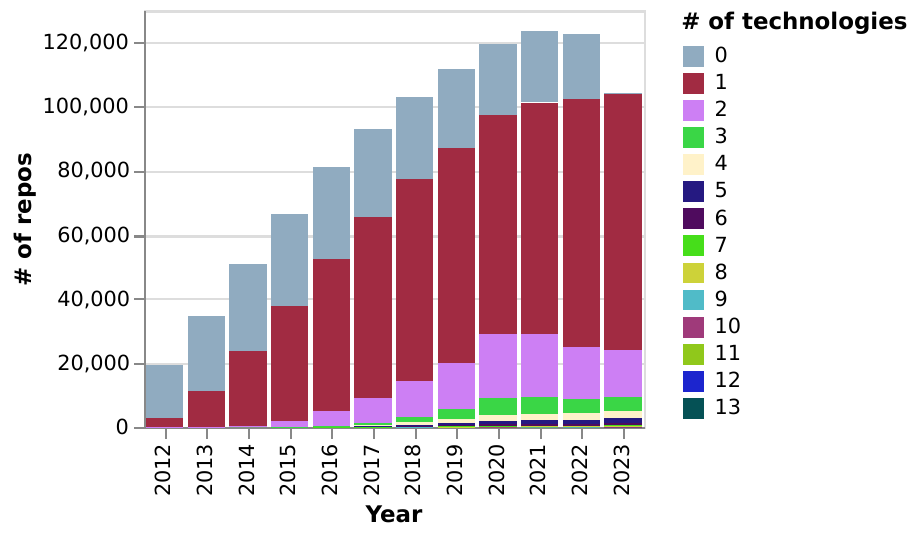}}
\caption{Number of repositories using at most a given number of technologies in a given year, by year.}
\label{fig:max_n_tools_counts}
\end{figure}

\begin{figure}[htbp]
\centerline{\includegraphics[width=\columnwidth]{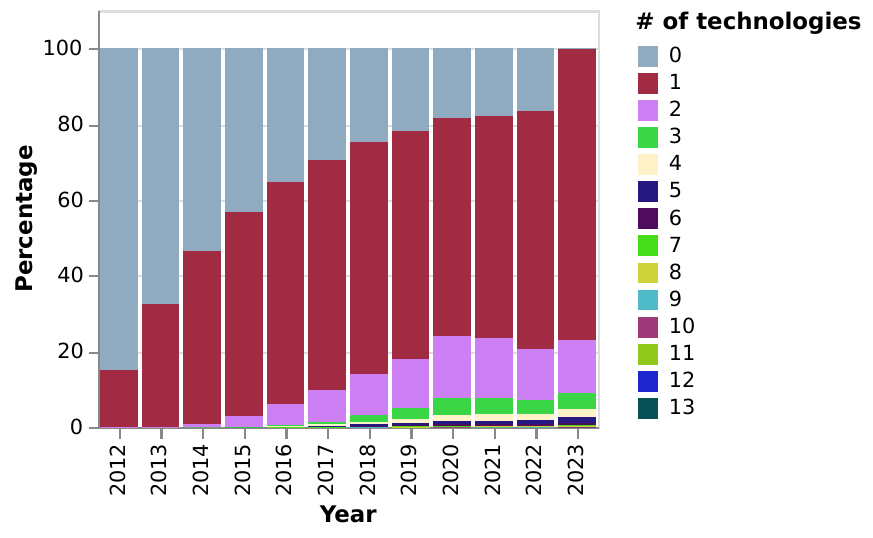}}
\caption{Percentage of repositories using at most a given number of technologies in a given year by year}
\label{fig:max_n_tools_percentages}
\end{figure}

\begin{figure}[htb]
\centerline{\includegraphics[width=.75\columnwidth]{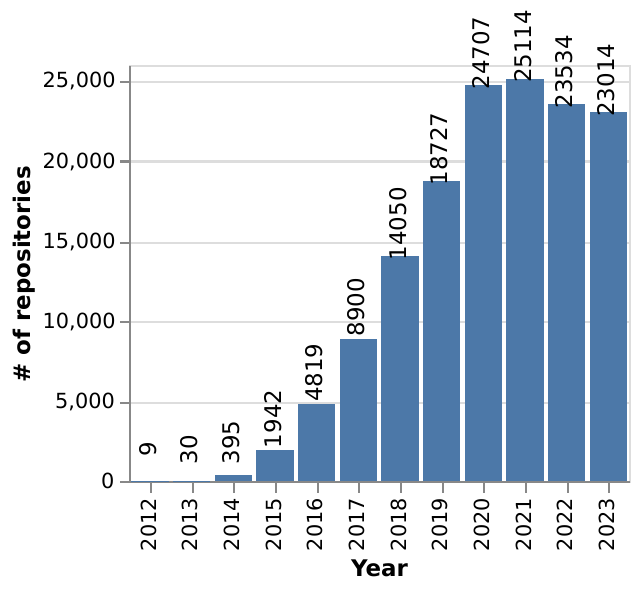}}
\caption{Number of repositories using more than one technology simultaneously in a given year, by year.}
\label{fig:repos_using_more_than_one_tool_per_year}
\end{figure}

In \cref{fig:max_n_tools_counts} and \cref{fig:max_n_tools_percentages}, we present the number and percentage of repositories using (at most) a certain number of technologies.
Some repositories have no single technology because, at a certain point (most likely, in their beginning), they did not use \ci.
In the first few years, one can see that the vast majority used just one technology, but that number quickly rose.
In the last 4 years, at least 23,000 repositories (per year) use more than one technology (\cref{fig:repos_using_more_than_one_tool_per_year}).
Note this is simultaneous usage, not different usage over time (which we will address in RQ3).
This finding agrees with both Mazrae et al. \cite{rostami_mazrae_usage_2023} and Goldazeh et al. \cite{golzadeh_rise_2022} studies of CI usage.

%\end{comment}

%Data also shows there is a significant number of repositories using more than one technology at a time (\cref{fig:max_n_tools_percentages}). This finding agrees with both Rostami Mazrae et al.'s \cite{rostami_mazrae_usage_2023} and Goldazeh et al.'s \cite{golzadeh_rise_2022} studies of CI usage.

\begin{comment}
{
  "2012": 19280,
  "2013": 34566,
  "2014": 50803,
  "2015": 66438,
  "2016": 81117,
  "2017": 93159,
  "2018": 102886,
  "2019": 111724,
  "2020": 119534, 24707/119534=20.7%
  "2021": 123431, 25114/123431=20.3%
  "2022": 122523, 23534/122523=19.2%
  "2023": 104353; 23014/104353=22.1%
}
\end{comment}

Although in the last few years there seems to be a decrease in the use of multiple technologies, clearly, there are many repositories still using them that the research community should try to support.

\begin{figure}[htbp]
\centerline{\includegraphics[width=\columnwidth]{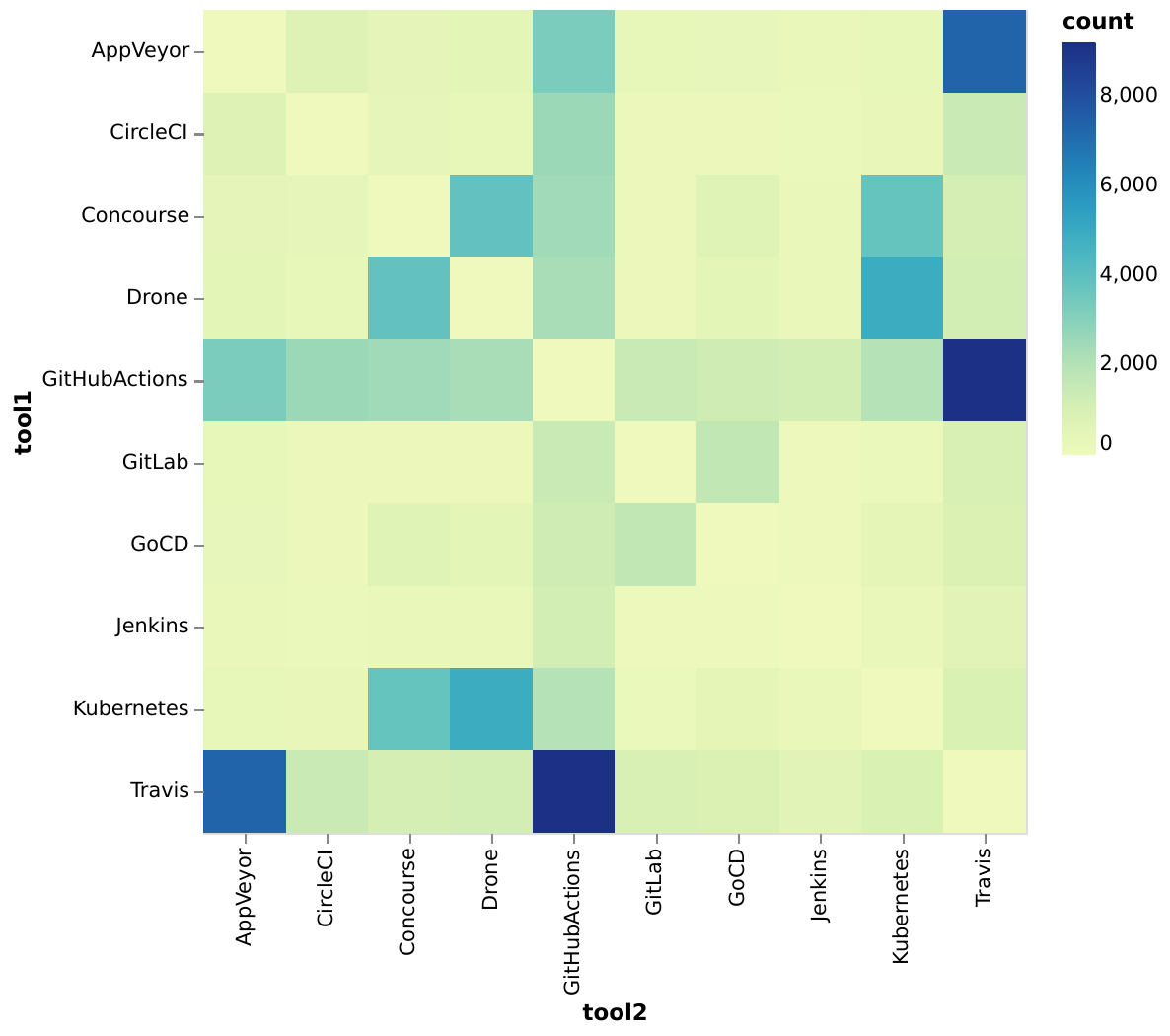}}
\caption{Heat map with the combination of technologies co-usage.}
\label{fig:maptooltool}
\end{figure}

In \cref{fig:maptooltool} we present a study of the co-usage of two technologies. 
Not surprisingly, GitHub Actions is used in combination with all the other technologies, but most significantly with Travis. Travis itself is quite used with AppVeyor and Drone with Kubernetes.

The stronger combinations deserve more attention as they represent more use cases and in particular more users the research community may help.

\begin{comment}

% appendix
\begin{figure}[!ht]
\centerline{\includegraphics[width=\columnwidth]{rq2/graphics/toolsperyear2ormore.pdf}}
\caption{Number of \ci tools used per repository per year, considering only the repositories with at least two tools.}
\label{fig:toolsperyear2ormore}
\end{figure}

%appendix
\begin{figure}[htbp]
\centerline{\includegraphics[width=\columnwidth]{rq2/graphics/maptooltoollesspopular.pdf}}
\caption{Heatmap with the combination of the (less popular) tools co-usage.}
\label{fig:maptooltoollesspopular}
\end{figure}

\end{comment}

Finally, in \cref{fig:plt}, we present the average number of \ci technologies used across different programming languages (we use the same set of languages as in RQ1).
As can be seen, all the projects written in these languages use, on average, more than one \ci technology.

\begin{figure}[!ht]
\centerline{\includegraphics[width=.9\columnwidth]{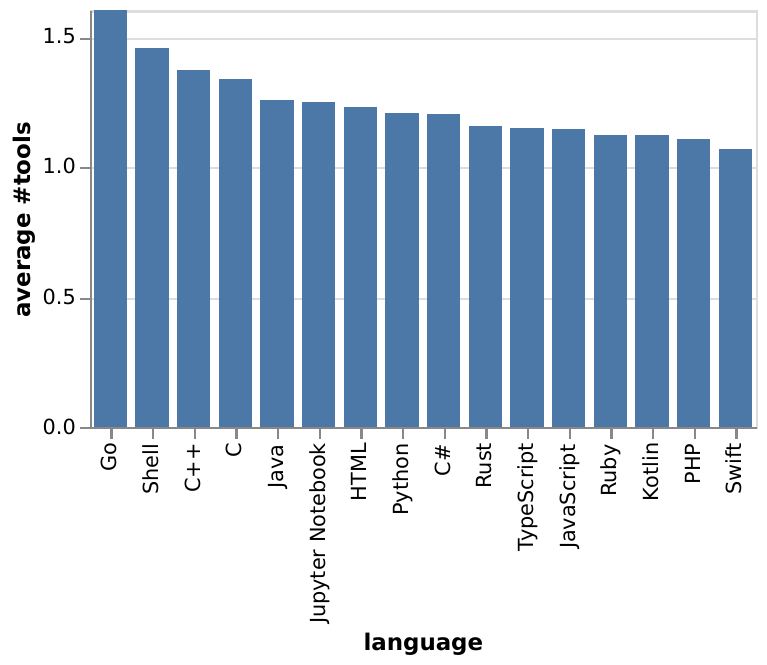}}
\caption{Average number of tools used per programming language repository.}
\label{fig:plas}
\end{figure}

This reinforces our previous conclusions: the usage of multiple technologies is quite present and deserves further study and support.

\section{RQ3: What is the evolution of CI/CD usage over time?} \label{sec:rq3}

With this RQ, we intend to understand the usage of \ci over the last 12 years.

\begin{figure}[!ht]
\centerline{\includegraphics[width=.9\columnwidth
]{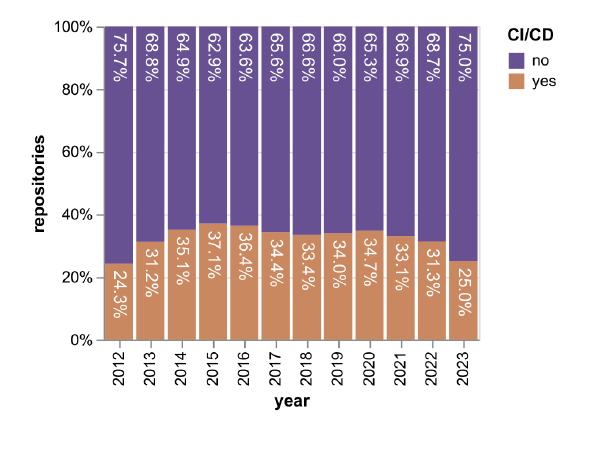}}
\caption{The usage of CI/CD per year.}
\label{fig:cicdperyear}
\end{figure}

We start by analyzing the usage of \ci starting in \st until now. For \cref{fig:cicdperyear}, we count the number of repositories created each year where, in their last commit, there is the presence of \ci technologies. As can be seen, even in repositories from 2012, almost one quarter included at least one technology; the percentage then increased until 2015 to more than 37\%, and has been since then in a gentle descending/stable curve. The drop in the last year is justified because many projects start without \ci, and only later, it is introduced.

\begin{figure}[!htb]
\centerline{\includegraphics[width=.75\columnwidth]{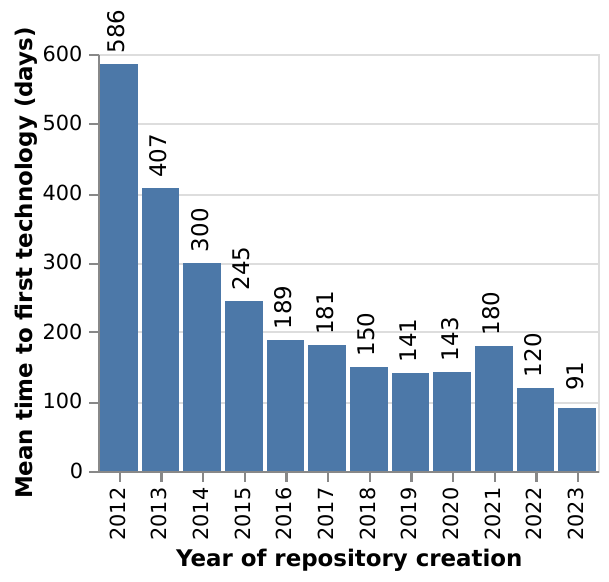}}
\caption{Mean time in days to first CI/CD technology detection by repository creation year.}
\label{fig:mean_time_to_first_tool}
\end{figure}

As seen in \cref{fig:mean_time_to_first_tool}, \ci technologies are being integrated into the development workflow sooner as time goes on. \ci's increasing popularity is due to older, possibly more complex, projects' adoption and new, perhaps simpler, projects that see value in continuous practices. The sharp decreases in 2022 and 2023 come from all analyzed repositories having at least one \ci technology in 2023. 

\begin{comment}
    [
      {
        "year": 2021,
        "mean_time_to_first_tool": 344.59956961730836
      },
      {
        "year": 2020,
        "mean_time_to_first_tool": 358.43326511422094
      },
      {
        "year": 2019,
        "mean_time_to_first_tool": 357.02274784970575
      },
      {
        "year": 2018,
        "mean_time_to_first_tool": 391.86376961545096
      },
      {
        "year": 2017,
        "mean_time_to_first_tool": 442.7954683951028
      },
      {
        "year": 2016,
        "mean_time_to_first_tool": 500.9927398989899
      },
      {
        "year": 2015,
        "mean_time_to_first_tool": 612.4228737180151
      },
      {
        "year": 2014,
        "mean_time_to_first_tool": 647.6275057509038
      },
      {
        "year": 2013,
        "mean_time_to_first_tool": 769.9350080237008
      },
      {
        "year": 2012,
        "mean_time_to_first_tool": 938.0151575271803
      }
    ]
\end{comment}

\subsection*{Usage of different technologies}

%To discover the changes in \ci technologies in the selected repositories over time, we plotted technology usage from 2012 until 2023. This is shown in \cref{fig:tool_counts}. 

\begin{figure}[!ht]
\centerline{\includegraphics[width=\columnwidth]{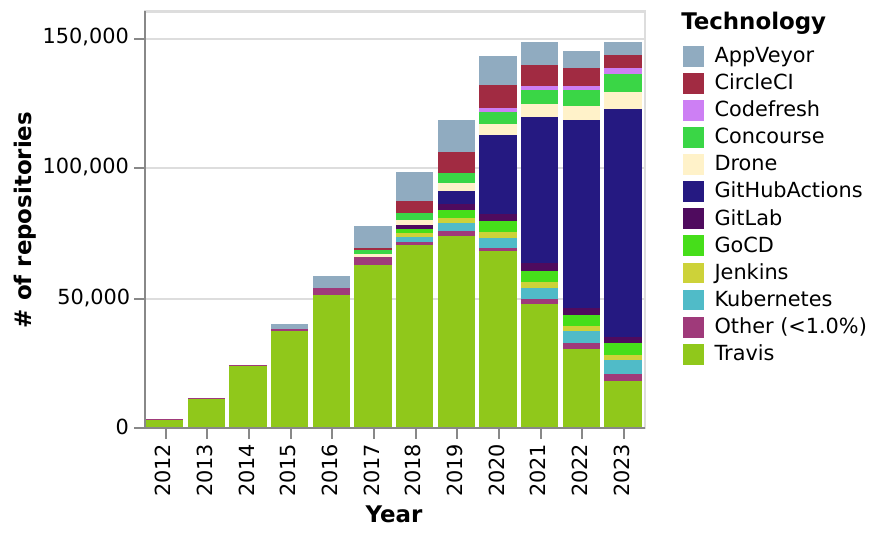}}
\caption{Number of repositories where each technology was detected by year.}
\label{fig:tool_counts}
\end{figure}

A repository may use more than one technology at a time, as seen in \cref{fig:tool_counts}. The technologies used in a repository in a given year are the union of the technologies of the snapshots retrieved in that year. In cases where a repository does not have any snapshots in a given year but has activity following that year, the technologies used for that year are the ones in the most recent snapshot up to that point in time, i.e., if a repository has snapshots in 2016 and 2018, but not in 2017, its used technologies in 2017 are the ones from the last snapshot from 2016. If a repository has no activity following a given year, its technology usage stops being considered. This methodology applies to all charts showing technology usage over time.

\cref{fig:tool_counts} shows two significant trends in CI/CD, Travis and GitHub Actions. Travis usage steadily increased from 2012 until it peaked in 2019 with 73284 repositories (36.6\%). Since 2019, Travis's usage has been declining. This coincides with the rapid adoption of GitHub Actions; from 2019 to 2020, there was a 502.8\% growth in the number of repositories using GitHub Actions, and from 2020 to 2021, there was an 86.6\% growth. Of the 36587 repositories that used Travis in 2019 and were still active in 2023, 59.7\% were using GitHub Actions and not Travis in 2023, 21.1\% used Travis and not GitHub Actions, and 16.9\% used both. Of the 87582 repositories using GitHub Actions in 2023, 45.4\% had no snapshots from before 2020. The exodus from Travis and the influx from newer repositories have been the main drivers for GitHub Actions growth. 

From this data, we can see a shift in the \ci technologies used 
over the years. This could mean developers tend to change their technologies and may need support for doing so. We thus investigate this shift further on.

\begin{figure}[!htb]
\centerline{\includegraphics[width=\columnwidth]{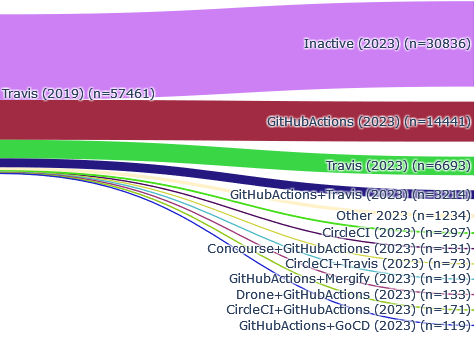}}
\caption{CI \ci technology stack transitions from repositories solely using Travis CI in 2019.}
\label{fig:sankey}
\end{figure}

\cref{fig:sankey} shows the top 10 CI/CD stack transitions from repositories that solely used Travis in 2019.
While many stop being active, 22.9\% moved from Travis to GitHub Actions. If we consider only the ones active, this
% 13924/(56423-30515)
53.7\% of all active Travis projects moved to GitHub Actions while.

From this information, we can conclude that developers actually change \ci technologies. However, there is little support for this kind of change. Thus, the research community has quite an interesting path to explore.

\subsection*{How often do projects change \ci technologies?}

\begin{figure}[htbp]
\centerline{\includegraphics[width=.7\columnwidth]{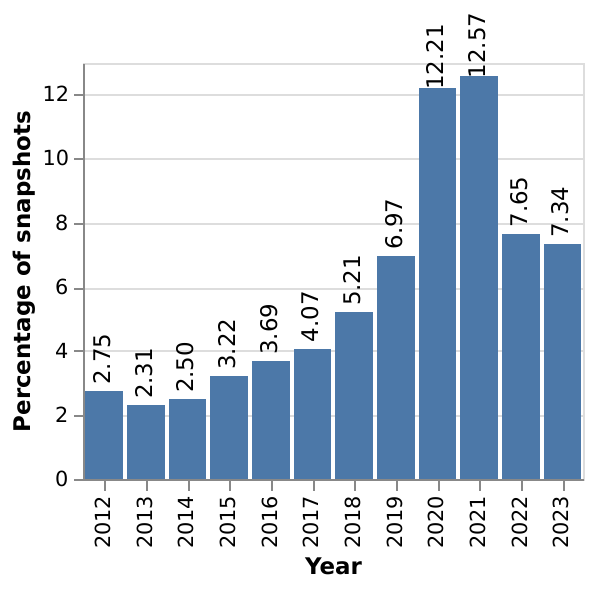}}
\caption{Percentage of snapshots with changes in the \ci technology stack from the previous snapshot by year (all repositories).}
\label{fig:tool_change_over_time}
\end{figure}

%\begin{sloppypar}
As \cref{fig:tool_change_over_time} shows, the percentage of snapshots with CI/CD changes compared to the previous snapshot grows steadily from 2013 (2.3\%) to 2019 (6.9\%) and peaks in 2020 (12.2\%) and 2021 (12.6\%), coinciding with GitHub Actions's explosive growth phase. Since 2021, this number has remained stable at almost 8\%. 
%\end{sloppypar}

\begin{comment}
    [
      {
        "year": 2019,
        "changes_percentage": 6.972504936958833
      },
      {
        "year": 2020,
        "changes_percentage": 12.214094972715662
      },
      {
        "year": 2021,
        "changes_percentage": 12.572072072072071
      },
      {
        "year": 2022,
        "changes_percentage": 7.654802744425386
      },
      {
        "year": 2023,
        "changes_percentage": 7.3390303341242875
      },
      {
        "year": 2018,
        "changes_percentage": 5.208388392621174
      },
      {
        "year": 2016,
        "changes_percentage": 3.685685254179881
      },
      {
        "year": 2017,
        "changes_percentage": 4.07309049110151
      },
      {
        "year": 2012,
        "changes_percentage": 2.750611246943765
      },
      {
        "year": 2013,
        "changes_percentage": 2.310458266391248
      },
      {
        "year": 2014,
        "changes_percentage": 2.4979480164158687
      },
      {
        "year": 2015,
        "changes_percentage": 3.2221800227876947
      }
    ]
\end{comment}

This is a very significant result since it shows that every year, between 2.3\% (in 2013) and 12.6\% (in 2021) of all snapshots include some kind of change in the \ci technologies used. This represents a very significant amount of technologies' shift, with all the known issues with that. Moreover, it is important to notice that this constant over time and there is no reason to think this may change in a near future. This means the research community can give a significant contribution to aid all these teams when they migrate and evolve their systems.

\begin{figure}[!htb]
\centerline{\includegraphics[width=.7\columnwidth]{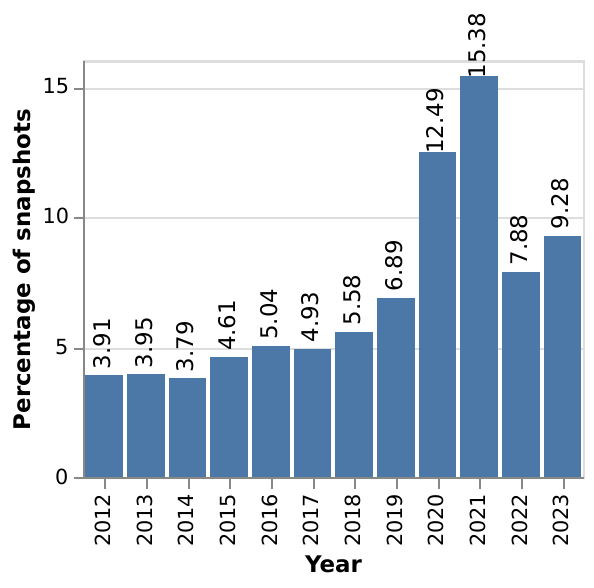}}
\caption{Percentage of snapshots with changes in the \ci technology stack from the previous snapshot by year for the set of repositories active from 2012 to 2023 (n=8296).}
\label{fig:tool_change_const_set_over_time}
\end{figure}

If we limit this analysis to a subset of active repositories from 2012 to 2023 (\cref{fig:tool_change_const_set_over_time}), we find similar trends but higher change percentages over time. For this analysis, we discarded all snapshots for each repository before the first, where we detected \ci technologies. This was done because we were not interested in a project's first choice of technologies. %Overall, projects tend to stick with their CI/CD \ci technology stack, and, looking at GitHub Actions and Travis's examples, most changes happen in fast paradigm shifts. 

\begin{comment}
    [
      {
        "year": 2018,
        "changes_percentage": 5.582274132600481
      },
      {
        "year": 2019,
        "changes_percentage": 6.894731284975188
      },
      {
        "year": 2020,
        "changes_percentage": 12.490047770700636
      },
      {
        "year": 2021,
        "changes_percentage": 15.384243128145567
      },
      {
        "year": 2022,
        "changes_percentage": 7.8821044546851
      },
      {
        "year": 2023,
        "changes_percentage": 9.276606280691619
      },
      {
        "year": 2014,
        "changes_percentage": 3.788982803847275
      },
      {
        "year": 2015,
        "changes_percentage": 4.605009633911368
      },
      {
        "year": 2016,
        "changes_percentage": 5.037202380952381
      },
      {
        "year": 2017,
        "changes_percentage": 4.9275919518134055
      },
      {
        "year": 2013,
        "changes_percentage": 3.946550652579242
      },
      {
        "year": 2012,
        "changes_percentage": 3.9080459770114944
      }
    ]
\end{comment}

The percentage of \ci technology stack changes is significant and is consistently higher in long-running projects, meaning projects continuously look for technologies that better fit their workflow. Although one could expect some stability over the years for mature projects, this is not the case as can been seen in the data. This reinforces the need for the research community to provide support for these changes.

\section{Discussion}\label{sec:discussion}

We have analyzed the 12 years of the usage of \ci. Answering RQ1, from the more than 600 thousand repositories collected, almost one third, 32.7\%, includes \ci technologies. This number is expected to grow as many projects start without the usage of \ci, which is added in later phases. We can also see one of the current technologies dominates the market, with 57.8\% of the projects using GitHub Actions. This number may be biased as we used GitHub as our source of projects, but, if this is the case, it means the source code manager has a significant impact in the choice of \ci technologies. 

Regarding the relationship with programming languages, it is possible to conclude some languages are more related to the usage of \ci than others. In particular, more recent languages, such as TypeScript, Go, or Rust have a higher percentage of projects with \ci than languages such as C, Java, or C\#. This may indicate a technological-generational gap between languages and \ci technologies, which should further be studied aiming to improve the adoption of \ci by older languages. 

Another interesting aspect is the fact that top-used languages such as R \cite{toplanguages} do not have projects that include \ci in a significant amount. This may indicate this kind of technology is not of easy adoption by less technically skilled users, as R is often used by researchers and mathematicians.

A surprising aspect we discovered during our work related to RQ2 is that many projects use more than one technology for \ci. In fact, up to 13 technologies may be used at the same time. In the last 4 years, each year, more than 23,000 projects include more than one technology, which accounts for about 20\% of all projects. From this, is possible to answer RQ2 negatively, that is, one technology is often not sufficient to cope with all the requirements of some projects. This raises several future work directions. A first path is to endeavor a deeper investigation is required to understand exactly why this happens. Moreover, it is necessary to investigate the interoperability of the technologies, how they work together, and how to provide the proper support for these users, as existing technologies tend to have tools (e.g., editors, interpreters) designed to be used independently of others.

In the analysis for RQ3, we can see about a third of the repositories have \ci, a number that is relatively stable over the years. However, we have shown the time between the creation of the repository and the adoption of \ci has been decreasing substantially.
Regarding the usage over the years, with RQ3 we discovered developers tend to change quite often \ci technologies. Indeed, there has been a massive change from Travis to GHA, but changes in technology are quite common. Indeed, in the last two years, more than 7\% of all snapshots include some kind of change in the \ci technological stack. If we consider long-run projects (projects running from 2012 up until now), this number is even greater. This means developers need support to be successful in these endeavors, which currently is mostly non-existent. This opens quite promising research paths, including techniques to support the evolution of this kind of artifacts (e.g., model-driven approaches, languages-based), but also for the human-computer interaction community on how to aid these developers being more effective and efficient.

\section{Threats to validity} \label{sec:threats}

There are multiple threats to the validity of our study, which we address in the following paragraphs.

Our study focuses on open-source software and, in particular, on projects hosted on GitHub.
Thus, our sampling does not include other kinds of software (e.g., propriety). Thus, we cannot generalize our conclusions to these other kinds of software projects. Nevertheless, many companies also have their software on GitHub, and one may expect workers from these companies to use similar technologies in other projects. Moreover, others have reached similar conclusions by interviewing developers \cite{golzadeh_rise_2022}.

Since we only used GitHub, we cannot say these results apply to projects in other code repository services. However, there is no reason to consider projects hosted on GitHub to be significantly different from other projects in other repository services.

Still regarding the use of GitHub as the source of the software projects we analyzed, we could observe a predominance of GitHub Actions. One of the main reasons for this may in fact be related to the use of GitHub as the source of projects. However, we also found GitLab Actions, the \ci technology used by another repository service (GitLab).

We collected our sample repositories by getting the 1,000 results sent by the GitHub API, doing it for every week in our time frame. This gave us more than 600,000 repositories, from which more than 200,000 have \ci. Although we could have collected more repositories, this would increase the time to retrieve them in a way that would make our work unfeasible. Moreover, the query did not impose any restriction on the results, except for the 10 starts we used to have some kind of ``quality'' metric for the projects. Thus, the repositories retrieved should not be biased in any other way.

We considered only technologies that we could identify through files in the repository. Indeed, from the 61 technologies identified by the Cloud Native Computing Foundation, we could not identify 14 technologies (plus 1 deprecated). Nevertheless, we were able to identify 64\% of all technologies.

Some technologies are detected through file contents and we cannot guarantee a random file would not have a certain string inside that matches. However, we defined content that would only make sense in the technology context, this probably did not happen. In any case, if it happened, was for a very small number of files that should not change the overall conclusions of our work.
%we actually had a problem with github actions that led to overdetection and we rechecked
  
We assumed that the presence of \ci artifacts (e.g., configuration files) means the underlying project is using such a technology. However, this may not be the case as some artifacts may be left forgotten from older usages.

\section{Conclusions} \label{sec:conclusions}

In this work, we have investigated more than 600,000 software projects, from which more than 200,000 include diverse \ci technologies.
We have characterized the current usage of \ci, related the technologies with programming languages, discovered that many projects use several technologies at the same time, and that projects tend to change their \ci technologies frequently. From this, several research paths can be seen, from better understanding why developers need to use several technologies simultaneously and to provide the proper support to them, to how to aid when developers want to evolve their technologies.

As future work, we plan to extend this work to other repository services such as GitLab, or Bitbucket. Another interesting research direction is to replicate this work in an industrial setting. We will also extend our work to consider the complete DevOps cycle.

%\section{Acknowledgments}
%We would like to thank Vasco Amaral and Alexandre TODO for the fruitful discussions about this paper.

%\begin{verbatim}
%???
%  \begin{acks}
%  ...
%  \end{acks}
%\end{verbatim}

\bibliographystyle{ACM-Reference-Format}
\bibliography{main}


\end{document}